\RequirePackage{fix-cm}
\documentclass[smallextended]{svjour3}       
\smartqed  
\usepackage{graphicx}
\usepackage[T1]{fontenc}
\usepackage[utf8]{inputenc}
\usepackage[dvips]{hyperref} 
\graphicspath{{./}{./figs/}}
\begin{document}

\title{Exploration in Free Word Association Networks: Models and Experiment
      }

\titlerunning{Exploration in Free Word Association Networks}        

\author{Guillermo A. Ludue\~{n}a$^\dagger$ \and Mehran Djalali~Behzad$^\dagger$ \and Claudius Gros}

\authorrunning{G.A. Ludue\~{n}a, M. Djalali~Behzad \& C. Gros} 

\institute{Guillermo A. Ludue\~{n}a$^\dagger$ \and Mehran Djalali~Behzad$^\dagger$ \and Claudius Gros \at 
           Institute for Theoretical Physics \\
           Goethe University Frankfurt \\ 
           Germany\\
			  $^\dagger$These authors contributed equally to this work.
		  }

\date{Received: date / Accepted: date}

\maketitle

\begin{abstract}
Free association is a task that requires a subject to express 
the first word to come to their mind when presented with a certain 
cue. It is a task which can be used to expose the basic mechanisms 
by which humans connect memories. In this work we have made use of 
a publicly available database of free associations to model the 
exploration of the averaged network of associations using a statistical 
and the \emph{ACT-R} model. We performed, in addition, an online experiment
asking participants to navigate the averaged network using their
individual preferences for word associations. We have investigated
the statistics of word repetitions in this guided association task.
We find that the considered models mimic some of the statistical 
properties, viz the probability of word repetitions, the distance 
between repetitions and the distribution of association chain lengths, 
of the experiment, with the \emph{ACT-R} model showing a particularly 
good fit to the experimental data for the more intricate properties as,
for instance, the ratio of repetitions per length of association
chains.

\end{abstract}


\section{Introduction}

Semantic memory~\cite{martin2001semantic}, which can be 
considered as a part of explicit memory, is responsible for 
the brain's ability to memorize the meaning of words and 
concepts and also their mental representation, including their 
properties and functions and the relation to each 
other~\cite{tulving1990priming}. One possible tool to 
study semantic memory is the task of free association, where 
a subject is asked to express the first word to come to mind 
related to some given cue. This task has a long history in 
psychology, dating back to the late 19th 
century~\cite{galtonFreeAssoc1880}. It is an instance of 
verbal fluency tasks which are commonly used for the study
of the structure of concept to concept associations within
the network organization of semantic memory \cite{goni2011semantic}.

A range of distinct semantic memory models have been 
suggested over the years, beginning in the sixties' and 
seventies' models recording dictionary information~\cite{quillian1967}.
It has been observed that priming effects, namely when a
semantically related cue has been presented to the test person
before, play a substantial role in memory retrieval and task 
performances~\cite{mcnamara_book,ratcliff-priming,tulving1990priming},
with prime and target possibly forming a compound 
object~\cite{ratcliff88}. 

A range of lexigraphical and associative semantic databases
have been collected over the years, like WordNet~\cite{wordnet}, 
the South Florida collection of free association, rhyme, 
and word fragment norms~\cite{NelsonNet} and
ConceptNet~\cite{conceptnet}. These word association networks 
typically exhibit small-world structures, with short 
average distances between words, together with strong local 
clustering~\cite{Steyvers-SemanticNetsModel,gros2011complex},
a property shared with lexigraphical spaces obtained
from word co-occurances~\cite{lund96}.

A comprehensive database of free associations, obtained 
from the participation of a large number of individuals 
(in the order of 6000), was made public by Nelson 
et al.~\cite{NelsonNet}.  This database, which we will
denote as {\em South Florida Free Associations}, SFFA,
in the following, can be considered an example
of a semantic space~\cite{steyvers2002}. The data essentially 
constitutes a weighted 
directed network, since both the forward and the backwards 
connectivity strength between all associatively related
pairs of around 5000 words, the vertices of the network, 
are provided. These association strengths are averaged over
all subjects taking place in compiling the database. Therefore,
individual associative preference may differ from that of 
the SFFA database. In addition, external effects like the 
environment, the last happenings before the experiment, 
etc. are ignored by the database. Also native and non-native
English speaker may have different associative preferences,
depending on the respective countries of origin.

In this work, we use the SFFA database as a basis 
for a guided association task. In this task, the subjects 
(either human or simulated models) navigate the network of 
words obtained from the SFFA database by connecting words
in a free association task. By comparing the statistical 
properties of word repetitions, as obtained by the the
associations chains created by human subjects, with those 
of the models for semantic memory retrieval, we expect to 
deepen our understanding of which properties may be important
for modelling semantic spreading~\cite{collins1975spreading}
on associative nets. Our works may be embedded in the context
of related studies employing the SFFA database, for which
the Google page rank has been computed and compared to the
experimental results of a lexical association task 
\cite{griffiths2007google}. It is also possible to 
simulated stochastic cognitive navigation on the SFFA
database in order to study possible mechanisms for 
information retrivial \cite{borge2010categorizing}.

Our work on exploration of free association networks can be 
considered also in the general context of semantic language 
networks \cite{sole2010language}, with the structure and
the dynamics of the respective network properties being 
studied intensively \cite{borge2010semantic}. From the
perspective of neurobiology an interesting question regards
the relation to possible underlying neural network
correlate for the association network studied here and
its relation to functional brain networks in 
general \cite{bressler2010large}. We also remark in this
context that the association network used for our study
corresponds to that of adults, with the development
of the human semantic network during childhood being
an interesting but separate topic \cite{beckage2011small}.

\section{Methods}

We set up an online experiment for a guided association task
\footnote{\url{http://itp.uni-frankfurt.de/\~mehran}}, attracting 
at total of 450 voluntary participants, mostly from the University 
of Frankfurt/Germany, the United Kingdom and the United States.
The goal of the experiment was to study associative exploration on
the SFFA network.

For the online experiment a randomly selected word from the SFFA, 
the cue, is presented to the subjects on the screen, along with 
a list of varying numbers of related words. The the list of words 
presented are all linked to the cue with a strength higher than 
5\% in the SFFA. The subjects are instructed to select the word 
from the list that seems to them most related to the cue. Then the 
selected word is taken as the next cue and presented to the subject
along with a new list of related words, extracted again from the 
SFFA. The subject can select one word, as in the previous step.
The task repeats itself until the subject voluntarily decides to
quit.

The sequence of words chosen by the participants is called a 
\emph{chain}, and the set of the 1688 chains collected constitutes 
the data from which statistical properties of the free association 
task were derived.

\subsection{Models}

In order to evaluate comparatively the data collected from the online
experiment we consider two models of memory retrieval. We use these
models to generate exploration chains in the SFFA network and
to compare the obtained simulated associative latching
with the actual data obtained from the online experiment.

\subsubsection{Mem model}
The \emph{Mem} model (for ``Memory'') consist in a random 
exploration through the SFFA network. The exploration starts 
at a random node (a word) from the network and moves to the 
next one, which is selected randomly with a probability 
proportional to the association strength to the present node,
as given by the SFFA database. The process is followed until 
a node is reached for which no outgoing link is provided. In 
addition to this simple exploration, there is a limitation for 
repeated words. When the exploration would jump to a word which 
was already visited during the exploration (the word is already in
the chain), it will be visited again only with a probability $c$. 
The word will therefore be ignored (at this step) with a probability 
$1-c$, which means that if such word is the only outgoing link of 
the present node, the exploration will end with said probability $1-c$.

The parameter $c$ is chosen in this work to the value $c=0.08$, for 
which it reproduces the experimental results for the distance between 
repetitions as closely as possible.

Notice that this model represents a memoryless probabilistic model for the
exploration of the word association network, in contrast to the one-step memory
model represented by the ACT-R model shown in next section.

\subsubsection{ACT-R model}
Within the \emph{ACT-R} model (Adaptive Control of Thought-Rational)
one tries to model both the activity and the retrieval dynamics of 
previously acquired memories~\cite{Anderson04anintegrated}.
In the \emph{ACT-R} model, a memory element $i$ has an activity 
$A_i(t)$, which is calculated as the sum of the base-level 
activity $B_i$ and an attentional weight $S_i$,
\begin{equation}
A_i(t)=B_i(t) + S_i(t) \; .
\label{eq:activity}
\end{equation}
The task attention term $S_i(t)$ is calculated as
\begin{equation}
S_i(t)=\sum_j \omega_j(t) W_{ji} \; ,
\end{equation}
where $\omega_j(t)$ is the attentional weight of the elements 
that are part of the current task, and $W_{ji}$ are the 
strengths of the connections between element $j$ and $i$.
For our purpose we have taken then $W_{ji}$ as the
association strengths of to the SFFA database. 

In our work, we have chosen to set $\omega_{j}(t)=1$  
if $j$ is the presently active memory (the node visited
at previous the moment), and $\omega_{j}(t)=0$ otherwise. 
Thus, in our version of the model, a word has a
higher task attention $S_i(t)$ if it is strongly related to 
the last observed word.

The base level activation $B_i(t)$ in Eq.~\ref{eq:activity} 
of node $i$ is given by
\begin{equation}
B_i(t) = \log\left( \sum_{t_k}^{t_k<t} \left(\frac{1}{t-t_k}\right)^{d}\right) \; ,
\end{equation}
where $t_k$ is the time of the $k$th last recall of the 
element $i$, and the exponent $d$ is a constant. Thus a 
given word has a high base activity level if it has been 
evoked many times lately.

Having defined the activity $A_i(t)$, the probability that an element $i$ is
remembered at time $t$, viz the retrieval probability, is given by
\begin{equation}
P_i(t) = \frac{1}{1+\exp\left[\frac{-(A_i(t)-\tau)}{s}\right]} \;
\end{equation}
where $\tau$ is the activity threshold and $s$ is a parameter 
introduced to account for the effect of noise onto the 
activation levels~\cite{Anderson04anintegrated}.
A word $i$ is recalled with probability $P_i(t)$ and the
averseness of a subject to repeat a word is given by $1-P_i(t)$.

Finally, the exploration of the network follows the same 
procedure as in the \emph{Mem} model. Being at site $j$
of the SFFA network a word $m$ is selected with probability 
$W_{mj}$ and accepted with probability $1-P_m(t)$. If this
word is accepted, than all $A_i(t)$, $B_i(t)$ and $S_i(t)$ are
updated.  If not, the procedure repeats until one word is selected
out of the list of candidates linked to the current site $j$.
The chain is terminated if all candidate sites are rejected.

For our simulations of this model, we have taken
$d=0.5$, $s=0.4$, $\tau = 0.35 * s$, which is a fairly standard
set of values~\cite{Anderson04anintegrated}.
A different set of values may be chosen to obtain a better fit to
the experimental results. However, it is our intention maintain a
range of values comparable with other studies in the literature.

\section{Results}

\begin{figure}[t]
\centering
\includegraphics[width=0.75\textwidth]{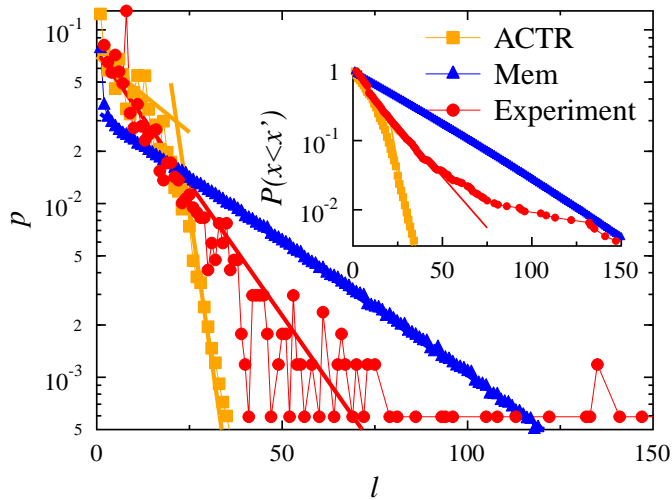}
\caption{Probability to observe word association
chains of length $l$, the vertical axis in log scale while the horizontal axis is linear. The data is obtained from the 1688 chains of the experimental data (red) and from $10^6$ chains generated by the
\emph{Mem} model (blue) as well as by the \emph{ACT-R} model (yellow).
The solid lines are respective exponential 
fits (see text for exponents). The inset shows the complementary cumulative distribution function of the same data, using the same representation. }
\label{fig:chainlengthfreq}
\end{figure}

In Fig.~\ref{fig:chainlengthfreq}, the probability distribution 
of chain lengths is shown in a normal-log representation, as well as
the corresponding complementary cumulative distribution function (CCDF) in the inset.
We observe an approximately exponential 
decay in the frequency of chain lengths for the experimental 
data as well as of both models. Also included in
Fig.~\ref{fig:chainlengthfreq} are exponential fits, given by
respective solid lines, evaluated using a maximum likelihood 
estimation (MLE)~\cite{newman-powerlawdistro,S}, evaluated with
the corresponding code from the \emph{GNU R} software package~\cite{R-book}.

The experimental data can be fitted well with a single exponential 
having an exponent $\lambda=-0.068(1)$. By chain length $\sim 50$ 
the number of data points is too low for reliable data analysis.
The \emph{Mem} model allows for larger chain length, having
an exponent $\lambda=-0.03593(3)$.

For chain lengths of size smaller than $\sim 20$ elements, 
the \emph{ACT-R} model follows closely the behavior 
of the experimental data, with an exponent $\lambda=-0.0400(1)$. 
There is a kink for chain lengths $\sim 20$, with larger
chain length becoming progressively more unlikely for the
\emph{ACT-R} model. This decay for larger chain lengths can 
be fitted well by an exponential with an exponent $\lambda=-0.335(2)$. 
The theoretical models' data has been obtained, for both the \emph{Mem}
and for the \emph{ACT-R} model, using $10^6$ chains generated from 
random starting points on the SFFA network. It is hence interesting, 
that the \emph{ACT-R} data show a substantial amount of scattering 
for small chain lengths.

The experimental data is scarce and noise for chain length $\sim50$
and longer, as only very few subjects enjoyed engaging in the task 
as long. One may hence disregard, for further data analysis, all 
long chains. This would, however, involve setting a somewhat arbitrary 
cutoff. We have tested this procedure and found that the property 
of the experimental data remains essentially unaffected when keeping 
or removing long chains. We therefore opted, for simplicity, to present 
the results corresponding to the whole sample, including long 
chains.

\begin{figure}[t]
\centering
\includegraphics[width=0.75\textwidth]{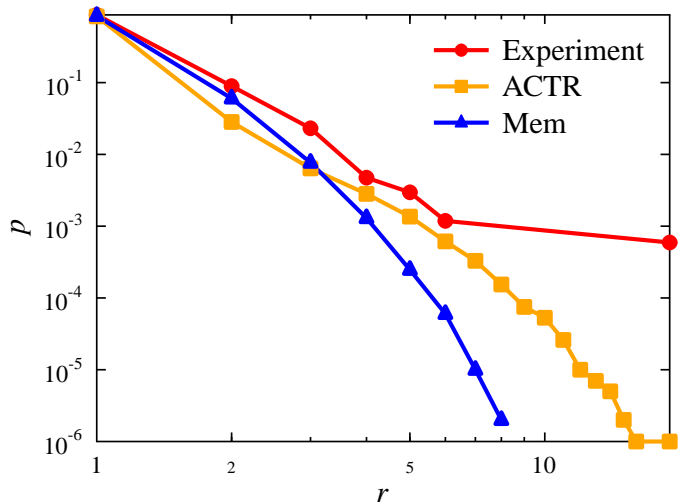}
\caption{The probability of observing $r$ word repetitions, averaged
over all chains lengths. The data is obtained from the 1688 chains of
the online experiment (red) and from the $10^6$ chains generated by
the \emph{ACT-R} model (yellow) as well as by the \emph{Mem} model (blue).}
\label{fig:repperchain}
\end{figure}

In Fig.~\ref{fig:repperchain} we present the probability $p$ that
a word is repeated one or more times, averaged over all chain
lengths. Only the data involving five or less repetitions is significant, 
for the results of the online experiment. The subject would prefer to 
stop a chain altogether and try with a new cue, than go on once a 
large number of repetitions did occur.
In this respect, we found that 19\% of all chains in our experimental results
end in a cycle.  We observe that the behavior of the chainlength distribution
remains unperturbed if these chains are not included.

The experimental results could, as a matter of principle, be
approximated by a power law, but the small number of data points 
does not allow for any definite judgement. This behavior seems 
to be shared with the \emph{ACT-R} models for the initial repetitions.
However, when the complete trend for larger number of repetitions is 
analyzed, a seemingly concave curve in the log-log plot can be
devised  both for the \emph{Mem} and for the \emph{ACT-R} model. This 
behavior cannot be cross-checked with the experimental data, due 
to the lack of data for larger numbers of word repetitions. We also 
tried to fit the data for the \emph{Mem} both with a Gaussian and 
with a simple exponential decay, but both approximations are not convincing.

\begin{figure}[t]
\centering
\includegraphics[width=0.75\textwidth]{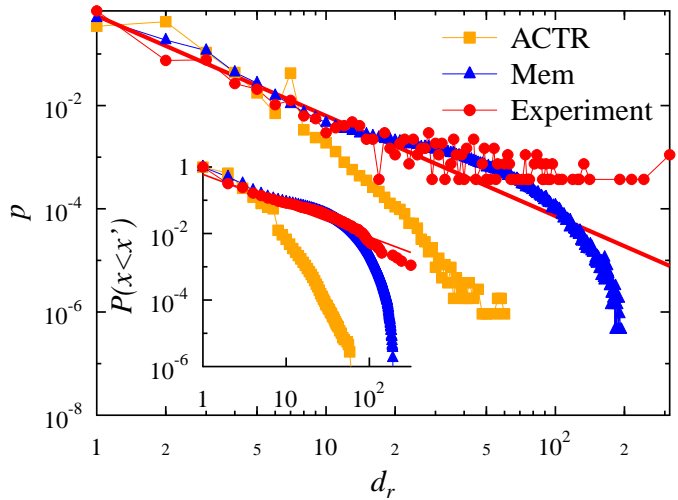}
\caption{Log-log plot of the distribution of distances $d_r$ between repetitions
of a given word. The data is obtained from the $10^6$ chains generated by the
\emph{ACT-R} model (yellow) and by the \emph{Mem} model (blue), as well
as from the 1688 chains of the online experiment (red).
The solid line represents, for comparison, a power law decay with
exponent $-1.9$. The inset shows the complementary cumulative distribution
functions of the same data, the solid line has in this case an exponent of $-0.9$. } 
\label{fig:distrepdistro}
\end{figure}

In Fig.~\ref{fig:distrepdistro} the distribution of distances 
between consecutive repetitions of the same word is presented.
All three datasets presented, for the two models and for
the experimental data, agree quite well up to repetition
distances of $\approx 10$. However, for larger repetition distances,
marked discrepancies are observed for both models, which
exhibit concave behaviors. The experimental data can, suggestively,
be approximated by a power-law with an exponent $\gamma=-1.9(1)$.

For the distribution of distances between repetitions, 
the \emph{Mem} model reproduces the experimental results 
somewhat better. This is not a coincidence, as the free
parameter of the \emph{Mem} model, the repetition probability 
$c=0.08$, has been selected to reproduce the experimental 
results for this property as closely as possible.

Although the decay of both models seem to fit relatively 
well the experimental data for small distances, they do 
not follow a similar law for the complete range. Due to 
the lack of enough data in the tail of the experimental
distribution, we do not consider this as strong evidence 
to disregard either of the models.

\begin{figure}[t]
\centering
\includegraphics[width=0.75\textwidth]{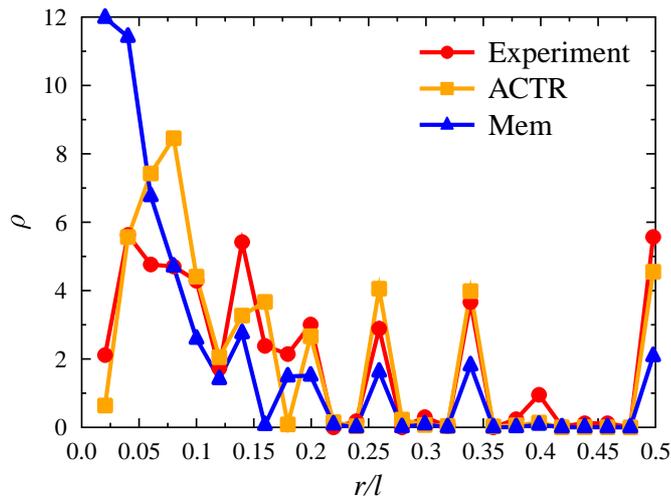}
\caption{The probability density $\rho$ to find a given ratio $r/l$ of
$r$ repetitions of a word per chain length $l$, obtained from the 1688 
chains of the experimental data (red) and from $10^6$ chains generated 
by the \emph{Mem} model (blue) as well as by the \emph{ACT-R} model (yellow).
The peaks at $l/r=2,3,4,\dots$ correspond to associative loops of length
$2,3,4,\dots$.} 
\label{fig:qrepchain}
\end{figure}

An interesting result can be observed in Fig.~\ref{fig:qrepchain}, where
we present the probability density $\rho$ to find a given ratio $r/l$ of
word repetitions ($r$) per chain length ($l$). A word that occurs three
times in a chain of length ten, to give an example, would
contribute to the frequency $\rho(r/l)$ of chains having a 
ratio of $r/l=3/10=0.3$. One observes a highly non-monotonic
distribution of ratios $r/l$. Experimentally
the maximal density is 0.5, which corresponds to a binary
loop like \emph{warm-cold-warm-cold-\dots}. There are additional
peaks at $r/l=1/3$ and $r/l=1/4$, corresponding to word repetition
loops of length three and four respectively.
It is evident from Fig.~\ref{fig:qrepchain}, that the \emph{ACT-R} model 
exhibits the same peaks as found by the online experiment with human subjects,
with approximately similar amplitudes for the respective word repetition
frequencies. This seems to be an indication that the \emph{ACT-R} model 
is suited for predicting the human behavior in this guided association task.
It may be also a hint that this distribution is strongly influenced by the
inclusion of a memory, which the Mem model lacks.

\begin{figure}[t]
\centering
\includegraphics[width=0.75\textwidth]{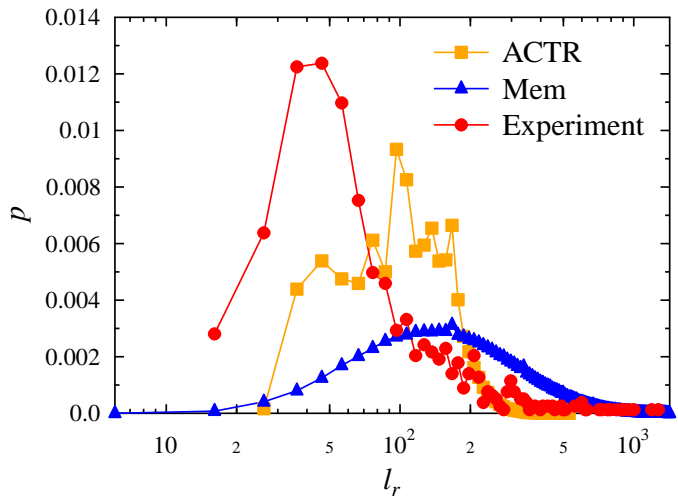}
\caption{The same data as in Fig.~\ref{fig:chainlengthfreq}, but
only for word association chains with a least one word occurring twice,
with histogram bin size 10,
and plotted in a normal-log representation.} 
\label{fig:probreplength}
\end{figure}

Finally we present in Fig.~\ref{fig:probreplength} the distribution 
(as an histogram) of chain length, just as in 
Fig.~\ref{fig:chainlengthfreq}, but retaining only word association 
chains with at least one repetition, which are mostly long chains. 
The human subjects tend to repeat words, on the average, substantially 
before both the \emph{Mem} and the \emph{ACT-R} model, which have have 
their distribution maxima at larger chains lengths. This result can
be regarded as robust, despite the observation that the results
from the online experiment is quite noisy. Note, however, that the
substantial scattering of the \emph{ACT-R}, which had been generated
using $10^6$ chain realizations, as for the other results.

\section{Conclusion}

Here we suggest, that online experiments for guided and related
associative tasks may provide interesting databases for human
association dynamics. The drawback of online experiments is,
to date, that there is no real control of how serious the individual
subjects take the task, some participants may just play around
randomly. There may be hence a certain fraction of non-characteristic
subjects which may, as a matter of principle, be taken into account 
by considering models with two populations of participants.
Our experimental database is however not large enough for this type
of analysis, for which a substantially larger number of participants
would be necessary. We however believe that this first online
experiment indicates that interesting data can be acquired. In particular
we analyzed the distribution of the lengths of guided associative world 
chains and various features of word repetitions. We attempted to
model the experimental results with cognitive models for human
memory retrieval dynamics, finding, in general, good qualitative
agreement.




\bibliographystyle{spmpsci}
\bibliography{freeassoc}

\end{document}